# The pressure dependence of the phonon spectra and elastic modulus of orthorhombic $GeSe$: the method of local density functional

F.M. Hashimzade, D.A. Huseinova, Z.A. Jahangirli, B.H. Mehdiyev[1], G.S. Orudzhev

Institute of Physics, National Academy of Sciences of Azerbaijan, AZ 1143, Baku, Azerbaijan

[1]E-mail: bachschi@yahoo.de

**Abstract.** The IR- and Raman-active phonon frequencies, as well as the elastic constants of orthorhombic $GeSe$, were calculatedas a function of hydrostatic pressure using the method of density functional in the ABINIT software package. Comparison with the published results of theoretical calculations and experimental data of the pressure dependence of Raman-active phonons has been carried out. Our calculations show that at a pressure of about 29 $GPa$ the crystal structure of $GeSe$ undergoes a continuous transition from simple orthorhombic to base-centered orthorhombic lattice.

PACS: 63.20.dk, 31.15A, 71.15.Dx

## 1.Introduction

Modern microelectronics based on the use of thin films grown on different substrates. The mismatch of the lattice constants results in a compressive or tensile stress in thin films. Furthermore biaxial and hydrostatic stresses a rise due to the difference of thermal expansion coefficients of the substrate and film. Under the influence of the applied pressure structural lattice parameters and electronic properties of crystals are substantially modified, and this should be considered in the development of various devices. Therefore, the study of influence of pressure on the structural, elastic and electronic parameters of the compounds is of great interest.

Earlier the pressure dependences of some of the phonon frequencies of the $GeSe$ compound have been investigated in [1,2], which also conducted experimental studies of the pressure dependences of the structural parameters and a low-frequency Raman-active shear mode $Ag$.

As a result of their research, the authors concluded that the density functional theory adequately describes not only the equilibrium structure and vibrational properties of highly anisotropic compound $GeSe$, but also effectively predicts details of pressure-induced changes of the lattice parameters and atomic positions therein.

The authors carried out a theoretical calculation of the pressure dependence of the interlayer shear mode $Ag$ in the approximation of the rigid layer. Their results show that this approximation gives a poor description of the pressure dependence of the interlayer mode.

The effect of hydrostatic pressure on the Raman spectra of this compound has been also studied empirically in [3]. The pressure dependence of Raman spectra $GeSe$ has been studied up to a pressure 7 $kbar$, and a large difference between the pressure coefficients of the low-frequency interlayer and intralayer modes of lattice vibrations was found.

Thus, the study of changes of interlayer and intralayer bonds in layered crystals, depending on the pressure requires more detailed study

In this paper we report a more detailed calculation of the effect of pressure on the IR - and Raman-active phonons, including the calculation of the elastic modules of the $GeSe$

## 2. Crystal structure and the method of calculation

It is known that the following four compounds from $A^4B^4$ group ($GeS$, $GeSe$, $SnS$, $SnSe$) have an orthorhombic lattice structure. The crystal structure is layered. Space group symmetry is $P_{nma}$ ($D_{2h}^{16}$) [4]. The positions of the atoms, in fractional coordinates, in the structure are follows: both atoms are at $4c$ as $\pm(x; 1/4; z)$ and $\pm(1/2-x; 3/4; 1/2+z)$.

The unit cell of the crystal contains eight atoms arranged in two layers, each consisting of four atomic planes in sequence: the cation-anion-anion-cation.

In this work, the calculations were performed first-principles density functional ($DFT-LDA$) using the basis of plane waves and pseudopotentials, implemented in a software package ABINIT [5].

In our calculations the exchange-correlation interaction was described in the local density approximation [6]. As a norm conserving pseudopotentials we used Hartvigsen-Goedekker-Hutter (HGH) pseudopotentials [7].

In the expansion of the wave function plane waves with a maximum kinetic energy up to 40 Hartree have been considered, which provides a good convergence of the total energy. Integration in the Brillouin zone (ZB) has been performed by using a partition of $4\times 4\times 4$ with a shift from the origin according to the Monkhorst-Pack scheme [8].

The lattice parameters and the equilibrium position of the atoms in the unit cell were determined from the condition of minimization of Hellmann-Feynman forces acting on the atoms. Minimization process was carried out until the force modules become less than $10^{-7}$ $Hartree/Bohr$. Then, interatomic force constants in the configuration space were calculated by the Fourier transform using ANADDB routine in the ABINIT software package [5]. These force constants were subsequently used to calculate the phonon modes at a number of arbitrary points of ZB.

Group-theoretical analysis predicts the presence of 12 Raman-active modes and 7 IR-active modes. In the Raman spectra the active modes are $A_g, B_{1g}, B_{2g}$ and $B_{3g}$, while in the IR spectra the modes appear with symmetry $B_{1u}, B_{1u}, B_{3u}$. $A_u$ mode is not active in the IR and Raman phonon spectra.

To obtain correct vibrational spectra of the crystals the calculations must be based on the equilibrium values of the lattice constant and the coordinates of the atoms therein. Therefore, it is necessary to perform an optimization of the structural parameters. For optimization of the crystal structure under normal conditions and under pressure BFGS algorithm was used to minimize for given stress tensor components $S_{ij}$.

The theoretical values of the structural parameters for GeSe at zero pressure and temperature are given in Table 1.

Considering the fact that the use of the LDA approximation usually leads to a small underestimation of the lattice parameters, the optimized and the experimental values of the lattice parameters are in a good agreement. This procedure was repeated for several magnitudes of pressure. At each pressure, before the calculations of phonon frequencies the structures were completely relaxed. The results are presented in Figures 1 and 2.

For comparison, these figures show the experimental results [1]. Figure 3 shows the pressure dependence of the phonon frequencies of the Brillouin zone center. Unfortunately, experimental pressure dependence of the phonon frequencies is known only for one mode $A_g$. This comparison is shown in Figure 4.

Figure 5 shows the pressure dependences of the elastic modules of the crystal $GeSe$.

## 3. Discussion

It can be seen that the theoretical dependence of the frequency of the phonon mode $A_g$ matches quite well the experimental results. Note, however, that our results are related to zero temperature and therefore the frequency is slightly higher than the frequency experimentally measured at room temperature. In addition, we compared the pressure coefficients for several modes, experimentally investigated in [3] at low pressures.

**Table 1.** Structural parameters for $GeSe$ at zero pressure and temperature.

| Parameters | $a(Å)$ | $b(Å)$ | $c(Å)$ | $x_A$ | $z_A$ | $x_B$ | $z_B$ |
|---|---|---|---|---|---|---|---|
| Theory | 4.305 | 3.76 | 10.568 | 0.1109 | 0.1173 | 0.4942 | 0.8550 |
| Experiment | 4.388[a] | 3.833[a] | 10.825[a] | 0.1115[a] | 0.1211[a] | 0.502[a] | 0.8534[a] |
| | 4.38[b] | 3.82[b] | 10.79[b] | 0.106[b] | 0.121[b] | 0.503[b] | 0.852[b] |
| | 4.381[c] | 3.834[c] | 10.847[c] | 0.110[c] | 0.124[c] | 0.504[c] | 0.844[c] |

[a]Ref.[4]
[b]Ref.[9]
[c]Ref.[2]

For low-frequency $A_g$ mode with a frequency of 39 $cm^{-1}$ we obtained a good agreement, namely, the experimental value of baric coefficient is 7, while the theory predicts 6.5. For mode with the frequency 174 $cm^{-1}$ the experimental value is 2.9, and the theoretical value is 2. However, with the modes of $A_g$ with frequency of 188 $cm^{-1}$ and $B_{3g}$ with frequency of 151 $cm^{-1}$, we observed, in contrast to the experimental situation, a negative value for baric coefficient. At present we do not have a convincing explanation of this difference. Fig.7 shows the Birch-Murnaghan equation of state [11], constructed for bulk modulus of 40.78 and the pressure derivative modulus of 5.26. The points on the curve are the values of the dimensionless total energy theoretically calculated by the LDF method. These results agree with the experimental values of [ 2]. We calculated the bulk modulus of elasticity in two different ways. First, just by optimizing the cell parameters we obtained the value of 40.7 $GPa$. Second, by using the calculated independent elastic modulus, we obtained the value of 36.8 $GPa$ .Both values are in a reasonable agreement with the experimental value of 379 kbar found in [2]. In this paper, to calculate the elastic constants of the $GeSe$ by method DFPT, which allows to determine the second derivatives of the total energy with respect to deformation and, thus, allows a direct calculation of the elastic constants. Orthorhombic crystals are characterized by nine independent elastic constants

On the other hand, we compared the values of the diagonal components of the tensor of the elastic modules with the values calculated from the slope of the longitudinal and transverse acoustic branches, and found that they are consistent with a maximum deviation of 4%. From Figure 1 we see that the pressure dependence of the relative compression for the direction perpendicular to the layers is unexpectedly low. This phenomenon was first pointed out by the authors of [1]. It is this fact that led the authors to conclude that the layering in the crystal GeSe is weak. The Young's modulus calculated for each crystallographic direction show that, indeed, the compressibility in the direction of the "c" perpendicular to the layers is less than in the direction along the crystallographic axis "a", although more than in the direction of the crystallographic axis b. The same "anomaly" has been observed for the longitudinal sound velocity in these directions. That is, the sound velocity in the direction perpendicular to the layers is higher than in the direction of "a" along the layer

( $v_a = 2733\ m/s, v_c = 3752\ m/s; v_b = 4141\ m/s$ ).

Calculations show that at a pressure of about 29 $GPa$ crystal structure undergoes a transition from a simple orthorhombic to base-centered orthorhombic. Apparently, this phase transition explains the jump in electrical resistance $GeSe$ observed at 250 $kbar$ in the experiment [10].

Fig.2 (a) shows the change in the internal parameters $x(Ge)$, $x(Se)$ with pressure. The sharp change in the $x(Ge)$, $x(Se)$ near 29 $GPa$ correlates with a sharp drop in the resistance at 250 $kbar$ (See inset in Fig.2(a)). The difference in the pressure is attributed to the effect of the temperature. Our calculations refer to zero temperature, while the experiments were carried out at room temperature. Raising the temperature, naturally, increases compressibility, which leads to a phase transition at slightly lower pressures.

## 4. Conclusion

The vibrational spectra of *GeSe* under pressure were calculated with ab initio technique and compared to the experimental results of [1,2,3] and the theoretical results of [1,2]. Our results for the phonon modes can be used to interpret further IK experiments in future research. The results are consistent with Raman experiments of [1,2,3]. We also study the pressure dependence of elastic properties and velocities of longitudinal and transversal sound waves of germanium selenide. Our calculations predict a possibility of a phase transition at about 29 $GPa$ from orthorhombic $P_{nma}$ ($D_{2h}^{16}$) to $C_{mcm}$ ($D_{2h}^{17}$): both atoms are at $4c$ as $\pm(0, 1/4, z)$.

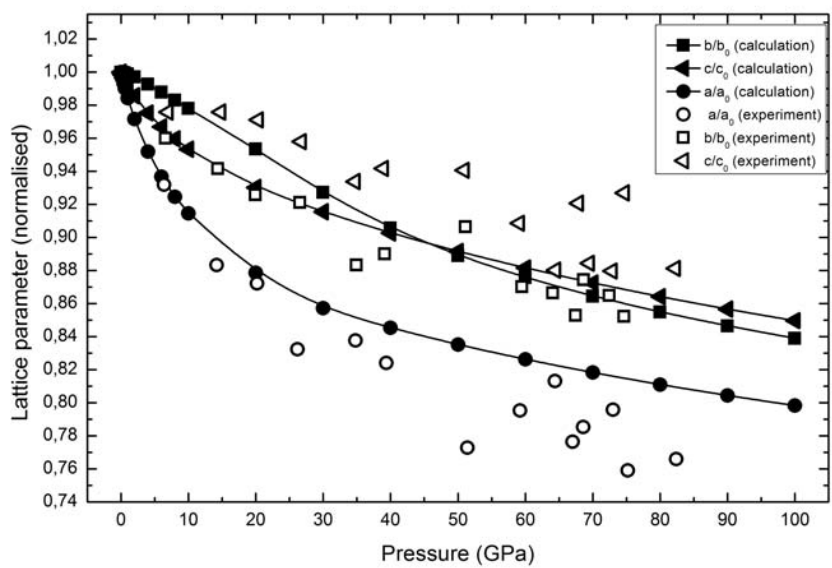

Figure 1. Normalized lattice parameters of $GeSe$ as a function of hydrostatic pressure. Solid symbols correspond to the calculated values and open symbols correspond to experimental data extracted from Ref.[10]. The solid lines correspond to the best fit to the calculated data.

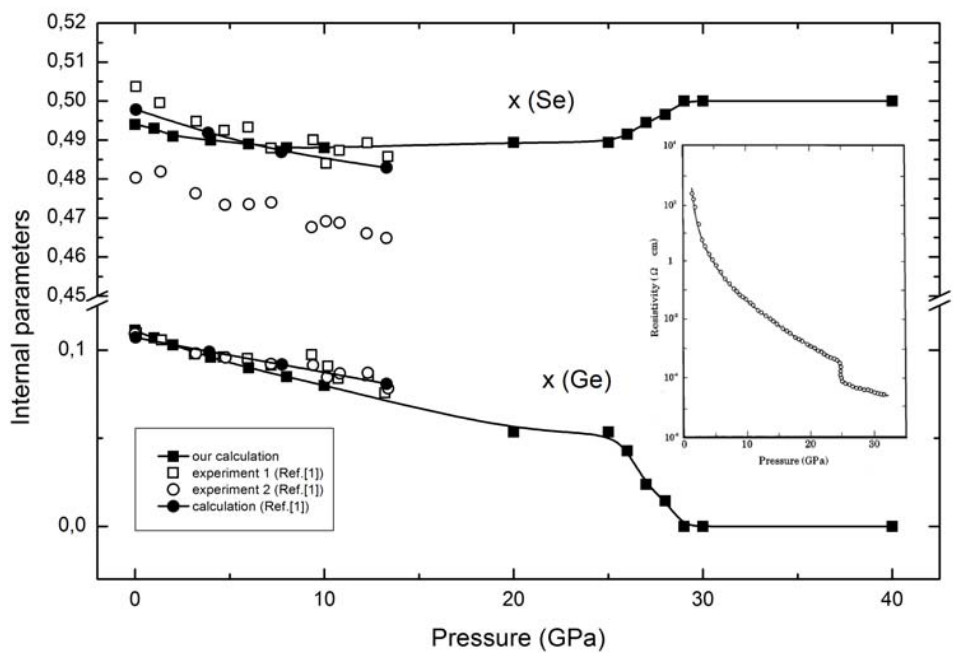

Figure 2 (a).

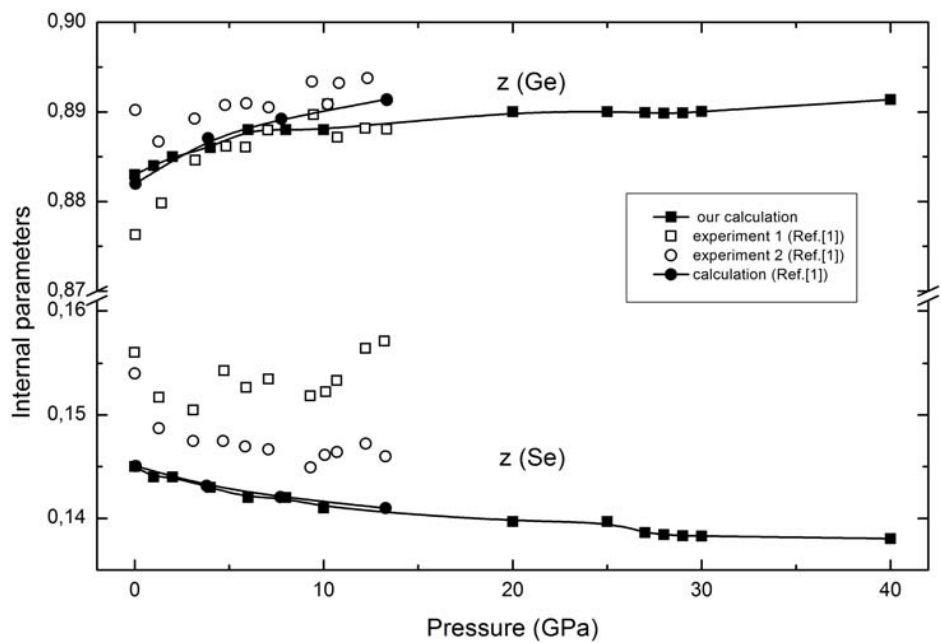

Figure 2 (b).

Figure 2 (a) and 2 (b). Pressure dependence of internal parameters. The inset of Fig. 2(a) shows the plot of electrical resistivity of $GeSe$ as a function of pressure extracted from Ref.[10]

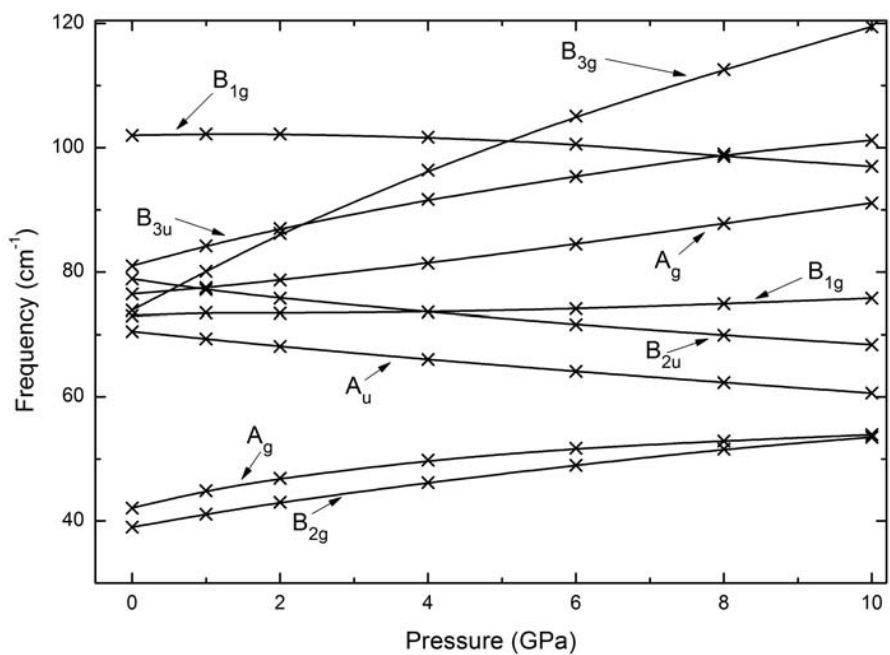

Figure 3 (a).

Figure 3 (b).

Figure 3 (a) and 3 (b). The pressure dependence of the phonon frequencies of the Brillouin zone center.

Figure 4. The hydrostatic pressure dependence of the $A_g$ shear phonon mode.

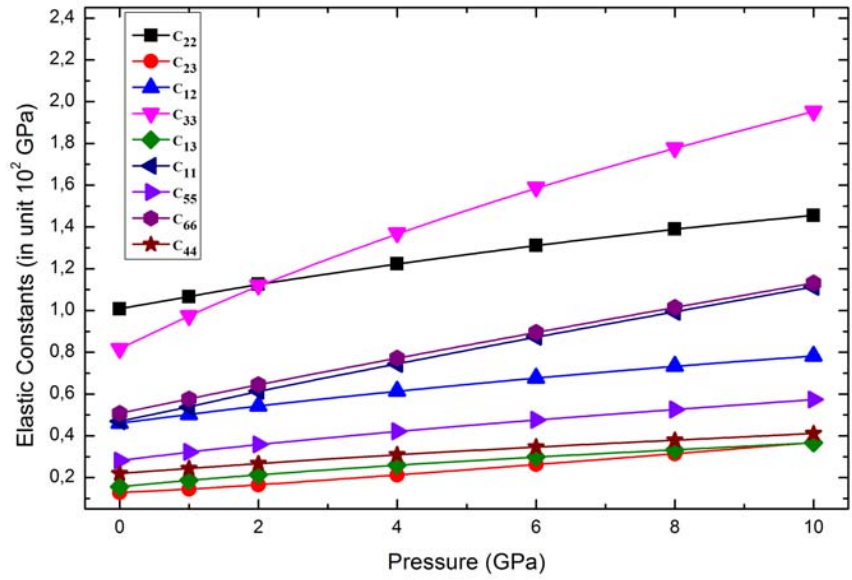

Figure 5. The pressure dependence of elastic constants.

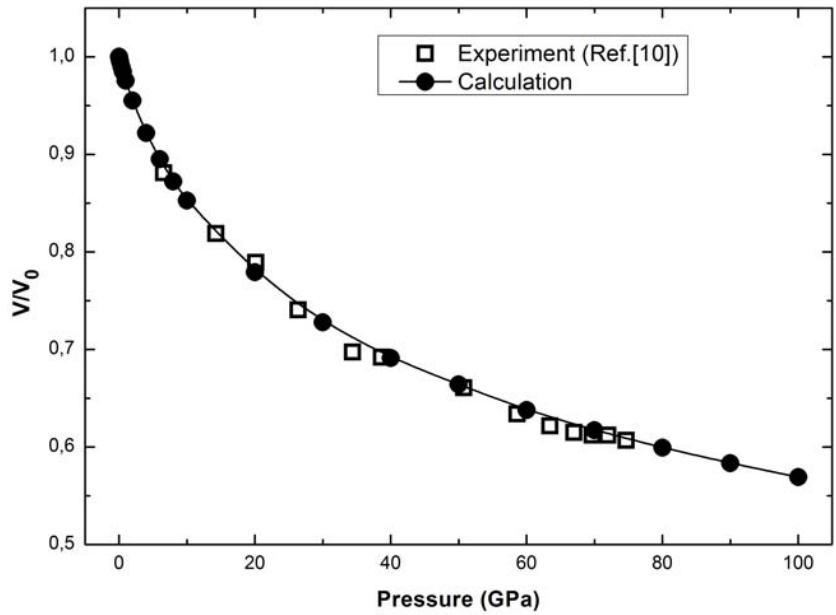

Figure 6. Volume versus pressure.

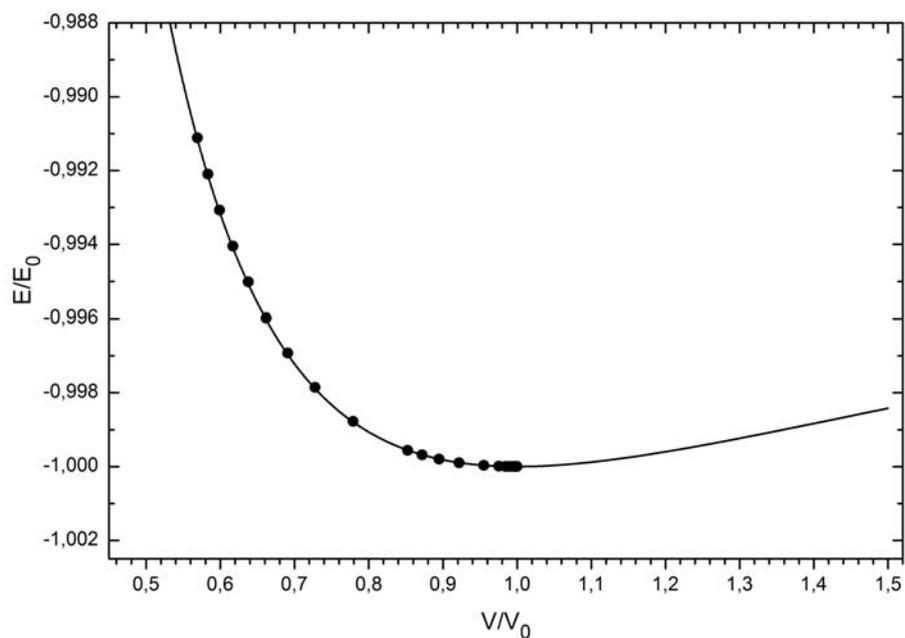

Figure 7. Pressure dependence of the volume of $GeSe$. Solid line corresponds to the fit to Birch-Murnaghan equation.

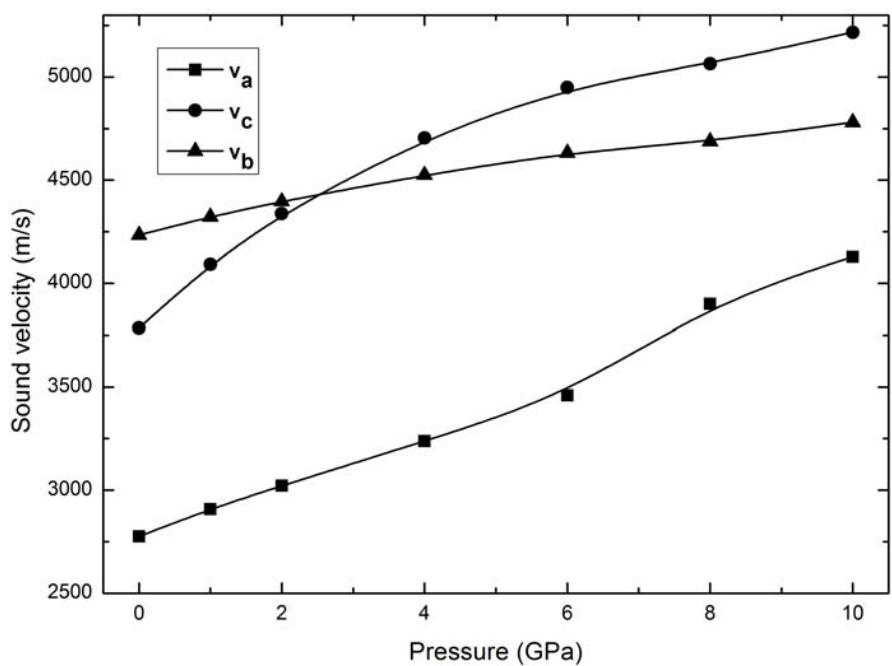

Figure 8. The pressure dependence of the sound velocity.